# Adaptive Push-Then-Pull Gossip Algorithm for Scale-free Networks

Ruchir Gupta, Abhijeet C. Maali, Yatindra Nath Singh, *Senior Member IEEE*
Department of Electrical Engineering, IIT, Kanpur
{*rgupta, abhicm, ynsingh*}@iitk.ac.in

*Abstract*—Real life networks are generally modelled as scale-free networks. Information diffusion in such networks in decentralised environment is a difficult and resource consuming affair. Gossip algorithms have come up as a good solution to this problem. In this paper, we have proposed Adaptive First-Push-Then-Pull gossip algorithm. We show that algorithm works with minimum cost when the transition round to switch from Adaptive Push to Adaptive Pull is close to $Round(log_2(N))$. Furthermore, we compare our algorithm with Push, Pull and First-Push-Then-Pull and show that the proposed algorithm is the most cost efficient in Scale-Free networks.

## I. Introduction

Real world networks are generally modelled as Scale-Free networks. Information diffusion in such networks is a resource consuming affair due to the distributed environment. Gossip algorithm seems to be a good solution to this problem. In gossip algorithm, in each round, a node randomly chooses one of its neighbours and exchanges information with it. We assume that information may start spreading in the network from any arbitrary node. Over a period of time, some nodes will get this information. Such nodes will be termed as informed nodes or *Spreaders* [7] as now they also take part in spreading the information further in the network by sharing it with their neighbours. Remaining nodes will be termed as uninformed nodes or *ignorant nodes* [7]. Terms *ignorant* and *spreader* are used to describe the status of a node. Gossip algorithms have different variants, some of them are described below.

**Push:** Spreader randomly chooses one of its neigh-bours and transmits the message.

**Pull:** Ignorant randomly chooses one of its neigh-bour and asks for the message.

**Push-Pull:** All nodes randomly choose a neighbour in each round. Both nodes become spreader if either of them is a spreader.

**Differential Push:** Spreader randomly chooses $k$ of its neighbours and transmits the message. Here $k$ is the ratio of node's degree and the average degree of its neighbours.

**First-Push-Then-Pull:** Nodes pushes information in initial few rounds and then switches to pulling information after these initial rounds [6].

Scale-Free networks have long tail degree distribution. We propose to classify the nodes into three types viz. hub or high degree nodes, low degree nodes and moderately high degree nodes. Though small in number, hubs are well connected in the network because of their high degrees. They play important role in the information dissemination as they can communicate with large group of neighbouring nodes. In such networks, moderately high degree nodes can play a crucial role in speeding up the information diffusion. We propose adaptive FPTP algorithm which involves above mentioned rate boosting nodes in the initial stages of information dissemination of information. Adaptive FPTP results in improvement over Push, Pull or FPTP algorithm in terms of cost and convergence time with high probability. We have suggested the optimum transition round to switch from Push to Pull in Adaptive FPTP algorithm and evaluated the number of rounds required for dissemination of a message to all nodes in the network.

The cost of the propagation per fresh spreader can be defined as the average number of calls or messages required to deliver a message to an ignorant node. For each round the cost is calculated as

$$Cost = \frac{Total\ number\ of\ spreader\ in\ the\ previous\ round}{Number\ of\ fresh\ spreaders\ in\ the\ present\ round}$$

The ignorant nodes that received message in current round and became spreaders are called *fresh spreaders*. In all graphs the cost is plotted as average transmissions/ ignorant node in each round of the dissemination. The minimum value of cost will be 1, when the number of spreaders doubles in next round. That means all the messages generated in the network in that round are delivered to ignorant nodes and no message get wasted. The cost becomes infinite when no ignorant receives the message in that round and the number of spreaders remains same as it was in last round. In this work, we have used PA model proposed by Barabasi *et.al.* [2] to generate the scale-free network. We have also assumed that cost for push and pull is same.

Rest of the paper is organized as follows. In Section 2, adaptive push and adaptive pull algorithms have been discussed. In Section 3, we present adaptive first push then pull algorithm. Section 4 concludes the paper.

## II. Adaptive Algorithms

Some adaptations in the existing Push and Pull algorithms can make FPTP algorithm more efficient in scale-free networks. In this section, we show that FPTP algorithm performs better when priorities are given on the basis of nodal degrees. Adaptive Push and Pull algorithms to be later used in First-Push-Then-Pull (FPTP) algorithm [6] for Scale Free networks are presented now.

Performance of push and pull algorithms has been improved by making them adaptive and consequently the performance of FPTP also get improved. In FPTP, both algorithms run back to back. As informed nodes spread the message, improved performance of the first one also boosts the rate of the later one. We have tested PUSH and PULL algorithms individually for Scale-Free Networks. The node in the network that generates information is termed as *initial spreader* or *source*. We carried out experiments by initiating algorithms with arbitrary initial spreaders. After averaging results from several runs, the performance measurement leads us to interesting results. These results formed the basis of Adaptive FPTP algorithm.

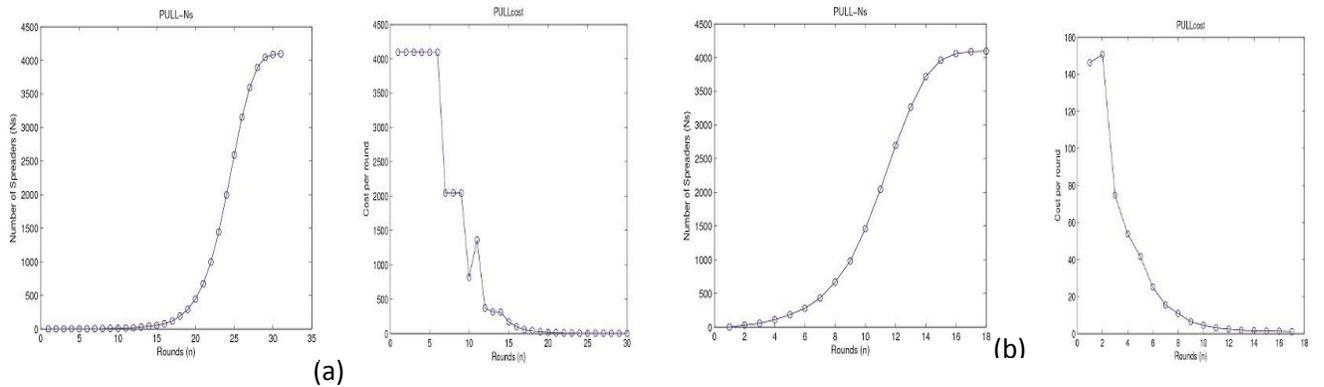

**Fig. 1** Message propagation (Number of spreaders) in Pull algorithm for $N = 2^{12}$ network for initiator with nodal degree (a) k=2 and (b) k=82.

Figure 1 shows the working of Pull algorithm for same parameters over same network. Initial spreaders are $k = 2$ and $k = 82$. The cost and the number of rounds for the complete dissemination of the message vary with the initial spreader degree. For higher degree initial spreader, cost and rounds are lesser when compared to the lower degree initial spreader case. This is because, according to Power Law degree distribution large number of lower degree nodes is connected to higher degree nodes, so in Pull algorithm there is high probability of lower degree ignorant nodes calling higher degree nodes. Initial spreader being a high degree node or hub, calls get positive response with high probability reducing the cost and increasing the rate. There is a probability of calling same hub by multiple ignorant neighbours. Hub forwards message to neighbours that asked for it but not to all ignorant neighbours. If all ignorant neighbours do not ask then hubs' capability get wasted. We remove this flaw in Adaptive Pull algorithm. Unlike Pull, Push algorithm is not much dependent on initial spreader degree. But we find small increase in number of spreaders for initial spreader degree in later rounds. The Pull algorithm (Figure 1) is quicker than Push algorithm. Push gives considerably longer tail i.e. it takes quite long in informing last few nodes which is a major disadvantage.

## A. Rich-neighbour

To retain the randomness of gossip we introduce Rich-neighbour method with a variable called Rich-neighbour. All the nodes in network maintain this variable. It points to the highest degree neighbour of a node. It will be maintained throughout the First-Push-Then-Pull strategy. Highly connected hubs can be the Rich-neighbours of multiple nodes which helps in controlled flooding in Adaptive Pull algorithm. In Scale-Free net-works, nodes are at a distance $d \leq 2$ from sufficiently high degree nodes with high probability [8]. Hubs become spreaders early by prioritization of Rich-neighbours in the algorithm.

## B. Adaptive Push Algorithm

The algorithm works in two phases, viz., deterministic and then randomized phase.

*1) Algorithm:* Algorithm 1 gives the algorithm for adaptive Push strategy with two phases explained be-low.

---

**Algorithm 1** Adaptive Push Algorithm

---

**if** $N(i)$ 2 *fresh spreader* — **then**
*Rich-neighbor message* $N(i)$
**else**
choose *neighbour* **randomly**, $N(j)$
$N(j)$ message $N(i)$

---

**Deterministic way:** When an ignorant node gets message it becomes spreader. We call a spreader node *fresh spreader* till it transmits its first message. So, a fresh spreader never randomly chooses neighbour. It gives first priority to the Rich-neighbour over randomly chosen neighbur and pushes the message. This might be called as making the gossip partially deterministic which helps to reach more number of higher degree spreaders at the earliest.

**Randomized way:** After the first push to its Rich-neighbour, a spreader does not remain a *fresh spreader*. Now onwards a spreader randomly chooses a neighbour and forwards the message.

Figure 2 shows the comparison of Normal and Adaptive Pull (2(a)) and Push (2(b)) algorithm for the network of 4096 nodes. Both algorithms are simulated on the same network.

Adaptive algorithm gives higher growth generating more number of spreaders. As hubs become spreader in the initial stages, Push phenomenon is enhanced.

## C. Adaptive Pull Algorithm

After transition from Adaptive Push, there is significant improvement in the following Pull algorithm. Normally, in PULL algorithm, an ignorant node randomly chooses a neighbour and asks for a message. After the Adaptive PUSH, there is more number of higher degree spreaders at the transition round, ignorant nodes can ask higher degree neighbour first. To take an extra advantage, we applied the similar logic while pulling the information. The algorithm works with same two phases as in adaptive push.

*1) Algorithm:* In this section we describe the Adaptive Pull algorithm. It will be explained in two phases. It works as given in algorithm 2.

---
**Algorithm 2** Adaptive Pull Algorithm
---

**if** $N(i) \in spreader$ — **then**
**stay quite** (No Pull only Push) **else** ($N(i) \in ignorant$ — **then**)
$N(i)$ ask for message Rich-neighbour
**if** Rich-neighbour $\in spreader$ — **then**
$N(i)$ message Rich-neighbour
**else**
choose neighbour ($N(j)$) randomly, in **next** round.
$N(i)$ ask for message $N($
**if** $N(j) \in spreader$ — **then**
$N(i)$ message $N(j)$

---

**Deterministic way:** As in Adaptive Push where fresh spreader pushes message to higher degree node, here, ignorant nodes pull or ask their higher degree neigh-bour (Rich-neighbour) first. As more number of higher degree spreaders are available after Adaptive push algorithm, ignorant nodes get positive responses from spreaders with high probability. Ignorant nodes give priority to Rich- neighbour in alternate rounds till they receive message.

**Randomized way:** After asking Rich-neighbour if a node does not get the message (if the Rich-neighbour is found to be an ignorant), it randomly chooses a neighbour and asks for a message. This phase retains the advantages of randomized gossiping such as simplicity and scalability. In Scale-Free networks, many lower degree nodes are connected to same high degree node which implies, a high degree node or a hub can be a Rich-neighbour of more than one of its neighbours. These neighbours call the same hub in one round and receive a message if the hub is in spreader state i.e. in Adaptive Pull algorithm, nodes may respond with message to multiple neighbours in a single round. This does not create redundant messages as the message will be transmitted only to those neighbours which are ignorant and have asked for it. A node can forward message to multiple neighbours in normal Pull also but all ignorant neighbours ask for the update randomly. This restricts the use of capability of hubs to communicate large number of nodes. Adaptive algorithm adds the advantages of flooding through hubs by utilizing the capability of hubs without creating redundant messages.

The important point about Adaptive Pull is that though it works slower at initial stages, its rate suddenly increases at some intermediate round e.g. $14^{th}$ round in Figure 2 (a). This sudden increase may come in earlier or in later rounds and depends on when hubs become spreader. Before that, ignorant nodes waste their calls to hubs in initial rounds because of dearth of spreader hubs. As hubs become spreader they speed up the rate. This implies that if spreader hubs are available in the initial round we get better results. The point to discuss Push and Pull algorithms individually is to show that how Adaptive Pull works better only when it is preceded by Adaptive Push. The earlier one helps to speed up the rate of later one by providing spreader hubs in the network. All the results are computed taking the initial source with least degree to check the worst case condition.

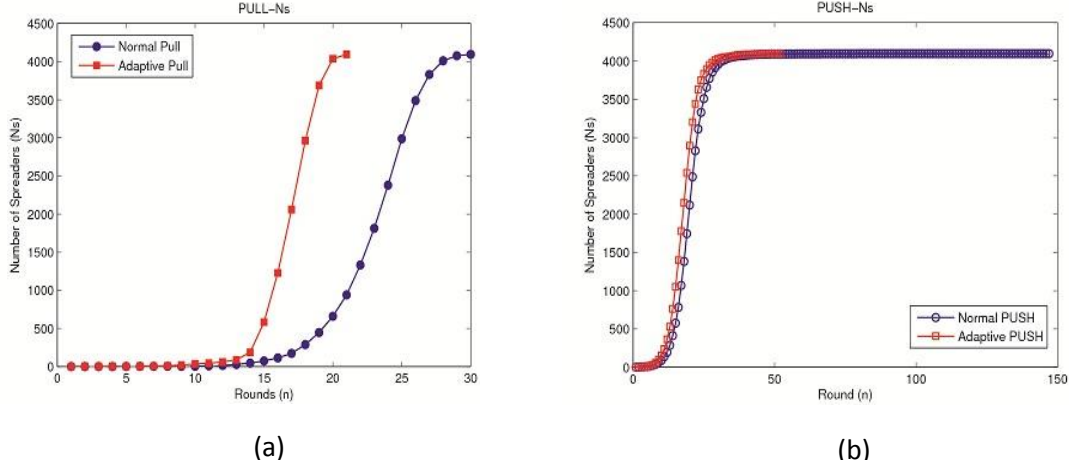

(a)                        (b)

**Fig.2: (a). Adaptive Pull vs. Normal Pull, (b). Adaptive Push vs. Normal Push**

### III. First-Push-Then-Pull Gossip Algorithm with Adaptive Push and Adaptive Pull

Switching from Push to Pull at different transition rounds gives different results. At optimum TR, the cost of the algorithm is minimum. The algorithm for adaptive FPTP is combination of Algorithm 1 and 2 by transitioning at optimum TR. The performance of Adaptive FPTP is better than Normal FPTP in both figures. We can observe sudden increase of rate in Adaptive FPTP than normal FPTP because the number of spreaders at the transition round is more in the former. Pull algorithm works as we explained in Section II-C2. From the transition round the difference between the two algorithms increases rapidly where Adaptive FPTP being the faster one and finishes earlier. This rapid growth is due to presence of spreader hubs at TR. One important point is that hubs are not the only nodes which become spreader with Rich-neighbour concept. Nodes having comparatively higher degree than the spreaders in the network also get benefited. The peaks in cost graphs are observed in both algorithms, but peak in Adaptive FPTP is quite less compared to normal FPTP algorithm.

Figure 3 shows the behaviour of two algorithms graphically. These simulation results are plotted. In figure, cost comparison is shown in first figure (3(a)) while the later one (3(b)) depicts comparison of rounds required. Both graphs are plotted against transition round (TR). The cost graphs are plotted on logarithmic scale. The minimum cost is obtained around $TR = log_2(N)$ with high probability. Here the value of N is taken as 32768.

$TR = 1$ means the algorithm switches to Pull in round one itself and so the dissemination is only through Pull algorithm. In Adaptive Pull, ignorant nodes waste their calls to Rich-neighbour, as with only Pull, most of the Rich-neighbours are in ignorant states in initial rounds. There is no improvement in cost in Adaptive Pull over normal Pull algorithm. But Rounds required are lesser in Adaptive case. This is because Adaptive Pull takes advantage of spreader hubs in the network in intermediate rounds.

Table I shows the simulation results of average cost and average number of rounds required. All the nu-merical values are calculated by averaging over 1000 runs. The cost and number of rounds required for Push and Pull are higher. We simulated for various transition rounds and tabled the values for $TR = log_2(N)$. Ali Saidi et al. [6] showed unlike the Push and Pull, the minimum average cost in FPTP remains nearly equal with small fluctuations and number of rounds increases with small difference. Adaptive FPTP algorithm retains these properties of FPTP.

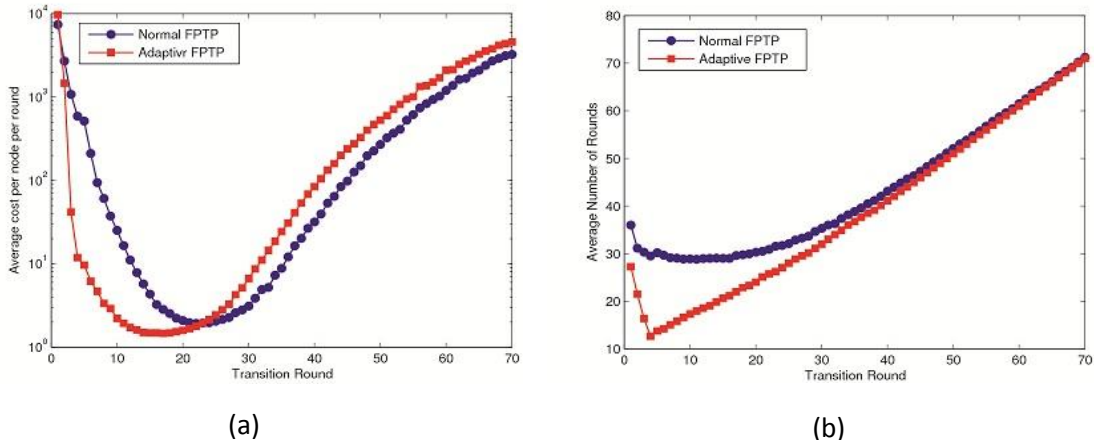

(a)                    (b)

**Fig.3: Comparison of normal FPTP and adaptive FPTP with respect to cost (a) and number of rounds (b).**

## V. Conclusion

In this paper, we have proposed adaptive first push then pull gossip algorithm for information diffusion in push and adaptive pull gossip algorithms based on the concept of rich neighbour for scale-free networks. Based on these results, we have proposed adaptive FPTP algorithm. Proposed algorithm performs better in terms of cost and convergence time in comparison with existing algorithms.

| *Number of Nodes | Push | | Pull | | FPTP | | Adaptive FPTP | |
|---|---|---|---|---|---|---|---|---|
| | Cost (avg) | Rounds (avg) | Cost (avg) | Rounds (avg) | Cost (avg) | Rounds (avg) | Cost (avg) | Rounds (avg) |
| 128 | 50.37 | 31.7 | 47.9 | 19.8 | 2.54 | 14.8 | 1.64 | 11.84 |
| 256 | 111.39 | 43.1 | 77.28 | 20.53 | 3.59 | 17.05 | 1.55 | 12.6 |
| 512 | 168.5 | 40 | 151.02 | 22.5 | 3.45 | 18.75 | 1.61 | 15.32 |
| 1024 | 384.12 | 52.4 | 299.55 | 26.33 | 3.33 | 20.4 | 1.61 | 16.65 |
| 2048 | 1023.8 | 75.56 | 630.6 | 29.3 | 3.14 | 21.75 | 1.59 | 17.35 |
| 4096 | 2263.1 | 97.8 | 1032.4 | 29.16 | 3.56 | 23.9 | 1.57 | 18.32 |

**Table 1: Average cost in transmission/node in each round and average number of round required**